\documentclass[final,1p,times]{elsarticle} 

\usepackage{graphicx}
\usepackage{amssymb} 
\usepackage{amsthm} 

\journal{Nuclear Physics A} 

\begin{document}

\begin{frontmatter} 

\title{Emission of Low Momentum Particles at Large Angles from Jet}


\author[label1,label2,label3]{Y.~Tachibana}
\author[label3]{T.~Hirano}
\address[label1]{Department of Physics, The University of Tokyo, Tokyo 113-0033, Japan}
\address[label2]{Theoretical Research Division, Nishina Center, RIKEN, Wako 351-0198, Japan}
\address[label3]{Department of Physics, Sophia University, Tokyo 102-8554, Japan}

\begin{abstract} 
We study dynamics of a QGP fluid induced by energetic partons propagating through it. 
We construct a (3+1)-dimensional QGP-fluid+Jet model.
When a jet traverses a uniform fluid, it 
induces a Mach cone structure of energy density distribution
and a vortex ring surrounding a path of the jet.
When a pair of jets travels through a radially expanding fluid,
 low momentum particles are dominantly induced at large angles from the quenched jet.
This result is qualitatively consistent with observation of the {CMS} Collaboration at {LHC}.
\end{abstract} 

\end{frontmatter} 


\section{Introduction}
Quark gluon plasma (QGP)
created in high-energy heavy-ion collisions behaves like a perfect fluid. 
Energetic partons created through initial hard scatterings
are subject to lose their energy during traversing the QGP medium, 
which is called jet quenching \cite{Bjorken:1982tu, Gyulassy:1990ye, Gyulassy:1993hr}.
Thus the jet quenching phenomenon is a promising tool 
to extract information about stopping power of the QGP. 
These partons could induce collective flow in the QGP by depositing energy and momentum. 
In fact, the CMS Collaboration observed a large number of low momentum hadrons
 at large angles from a quenched jet \cite{Chatrchyan:2011sx}.
The sum of these low momentum hadrons together with a quenched jet
balances
that of the momentum of a jet 
created at the same position and 
propagating in the opposite direction. 
Therefore the result can be interpreted as manifestation of a wake of the QGP fluid 
by the jet quenching.

In this paper, we construct a relativistic hydrodynamic model to 
demonstrate a response of the QGP fluid to jet propagation through it.
We model a source term 
in relativistic hydrodynamic equations which exhibits 
deposit of the energy and momentum from traversing jets. 
Without linearisation we solve these non-linear hydrodynamic equations 
numerically in full (3+1)-dimensional space 
and describe responses of expanding QGP medium to propagation of energetic jets.

\section{QGP-Fluid+Jet Model}
\label{}
To study the dynamics of the QGP fluid induced by jets, we use relativistic hydrodynamic equations with source terms
\begin{eqnarray}
\partial_{\mu}T^{\mu \nu}(x)=J^{\nu}(x).\label{eqn:hydro_source}
\end{eqnarray}
Here $T^{\mu \nu}$ is the energy-momentum tensor of the QGP perfect fluid and $J^{\nu}$ is the density of the four momentum deposited from the traversing jets.
We employ massless ideal gas equation of state, $P =e/3$, where $e$ and $P$ are energy density and pressure, respectively.
We assume that the deposited energy and momentum
are instantaneously thermalised inside a fluid cell.
For simplicity, we consider collisional energy loss only.
When a massless jet partcle travels through the QGP fluid, the source term is given by 
\begin{eqnarray}
J^{0}(x)= -\frac{dp_{\rm jet}^0}{dt}  
\delta^{(3)}\left(\mbox{\boldmath $x$}-\mbox{\boldmath $x$}_{\rm jet}(t)\right),\:
\mbox{\boldmath$J$}(x)=\frac{\mbox{\boldmath$p$}_{\rm jet}}{p_{\rm jet}^0}J^{0}(x).
\end{eqnarray}
We solve Eq.~$($\ref{eqn:hydro_source}$)$ numerically without linearisation for a given initial condition.
In this way, we describe collective flow induced by jets on top of
an expanding background QGP fluid in high-energy heavy-ion collisions.
Particle distributions can be calculated using Cooper-Frye formula \cite{Cooper:1974mv}
in an isochronous hypersurface in the calculation frame.

\section{Results}
\label{}
We consider two different situations.
As a test case, we consider one jet particle traveling through a uniform fluid.
As a more realistic case, we consider a pair of jets traveling 
in opposite directions through an {\it expanding} fluid.
We assume that jet particles are massless and travel in a straight line. 

\subsection{One jet traveling through a uniform fluid}
\begin{figure}[htbp]
    \begin{center}
      \includegraphics[width=100mm]{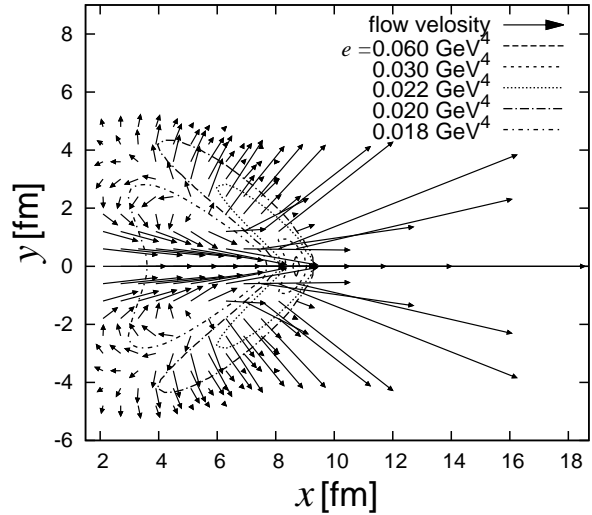}
    \end{center}
  \caption{Contours of the energy density distribution and the flow velocity field of the fluid
at $t=9$ fm/$c$.} 
  \label{fig:Mach}
\end{figure}

Figure \ref{fig:Mach} 
shows the energy density distribution and the flow velocity field
of the QGP fluid in a plane including a jet axis at a fixed time $t=9$ fm/$c$.
Initial energy density of the fluid is $0.019\:{\rm GeV^4}$.
The jet particle starts its propagation in the $x$-direction 
at the speed of light from the origin at $t=0$ fm/$c$.
One sees the so-called Mach cone, namely a ``V-shaped" cone-like structure
of higher energy density region in three dimensional space. 
The jet particle is followed by a strong collective flow in the positive $x$ direction, which leads to
lower energy density behind the cone than that of the backgrounds.
Notice here that the background energy density is $e=0.019\:{\rm GeV^4}$.
On the surface of the cone, flow velocity is nearly perpendicular to the surface.
Behind the cone, one sees vortices around $(x, y)=(5$ fm, $\pm 2$ fm) which form
a vortex ring around the path of the jet in three dimensional space.

\subsection{A pair of jets traveling through an expanding fluid}
Next, we consider a pair of jets 
traveling through a QGP fluid which expands radially.
Initial energy density distribution is
\begin{eqnarray}
e(t=0,\:\mbox{\boldmath $x$}) =
\left\{
\begin{array}{cc}
\frac{3T_0^4}{\pi^2} \exp \Big(-\frac{\mbox{\boldmath $x$}^2}{2\sigma^2} \Big),\:&(|\mbox{\boldmath $x$}|\leq R)\\
0.\:&(|\mbox{\boldmath $x$}|>R)
\end{array}
\right.
\end{eqnarray}
Here, we choose $T_0=0.5$ GeV, $\sigma=1.5$ fm and $R=6$ fm.
Without jets, this fluid expands isotropically.
A pair of jets
is supposed to be created at $\left( x, y, z \right) =
\left(3\:{\rm fm},\:0\:{\rm fm},\:0\:{\rm fm} \right)$
with the common initial energy $150\:{\rm GeV}$, both of which
are traveling in the opposite direction.
One jet goes in the positive $x$ direction and is observed as a leading jet.
The other goes in the negative $x$ direction,
 loses more energy and is observed as a sub-leading jet.

To see the momentum acquired by the QGP fluid from the jets,
we calculate total momentum distribution
as a function of polar angle measured from the leading jet axis
subtracted by the distribution in the case without jet propagation:
\begin{eqnarray}
\label{eq:mom-dist}
\Delta \left( \frac{ d \langle p^{||}\rangle _i }{d\cos \theta}\right) \equiv \sum_{p\in i}\frac{dp^{||}}{d\cos \theta} - \sum_{p\in i} \left. \frac{dp^{||}}{d\cos \theta}\right|_{\mbox{\tiny no jet}},
\:i={\mbox{high-}}p,\:{\mbox{mid-}}p,\:{\mbox{low-}}p,
\end{eqnarray}
where $\theta$ is the polar angle,
$p$ the momentum of the particle from the fluid and $i$ the index of the momentum regions.
Summation of the particles is taken over in the three different momentum regions
such as ``$\mbox{high-}p$" ($p>8\:{\rm GeV}$), 
``$\mbox{mid-}p$" ($4\:{\rm GeV}<p<8\:{\rm GeV}$) and 
``$\mbox{low-}p$" ($0\:{\rm GeV}<p<4\:{\rm GeV}$). 
We calculate Eq.~(\ref{eq:mom-dist})
 through the Cooper-Frye formula and do not add final jets' momentum:
\begin{eqnarray}
\sum_{p\in i}\frac{dp^{||}}{d\cos \theta}=\int_{p\in i} \left(p^0\right)^2 dp^0 d\phi \left(p^0\cos\theta\right)\frac{dN}{d^3p},
\end{eqnarray}
\begin{figure}[htbp]
   \begin{center}
     \includegraphics[width=80mm]{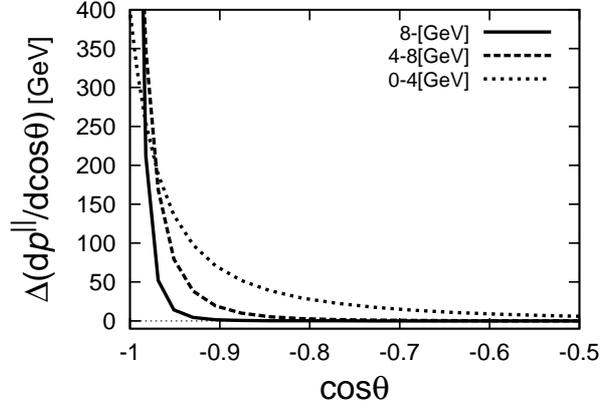}
   \end{center}
 \caption{The angular distribution of the fluid's momentum received from the jets.
Solid, dashed and dotted lines correspond to three different momentum regions,
``$\mbox{high-}p$" ($p>8\:{\rm GeV}$), 
``$\mbox{mid-}p$" ($4\:{\rm GeV}<p<8\:{\rm GeV}$) and 
``$\mbox{low-}p$" ($0\:{\rm GeV}<p<4\:{\rm GeV}$), respectively. }
 \label{fig:momentum}
\end{figure}
Figure 
\ref{fig:momentum} shows $\Delta \left( \frac{ d \langle p^{||}\rangle}{d\cos \theta}\right)$ as a function of $\cos \theta$
for three different momentum regions. 
Note that $\cos\theta=-1$ corresponds to the direction of the sub-leading jet.
A single prominent peak appears in this direction for each momentum region and
no additional peak away from the direction of the sub-leading jet, which is expected from the Mach-cone structure, is seen.
At large angles from the sub-leading jet $(\cos\theta>-0.94)$, low momentum particles are dominant. 
This result is qualitatively consistent with the {CMS} data \cite{Chatrchyan:2011sx} and indicates
importance of medium responses to jet propagation to understand
the modification of jet structure.

\section{Summary}
\label{}
In this paper, we studied the dynamics of the QGP fluid induced by jet propagation. 
We constructed a {QGP}-fluid+Jet model and performed two simulations.
In the case of  one jet traveling through a uniform fluid, a Mach cone appears
in the energy density distribution
and a vortex ring is formed around the path of the jet behind the cone.
In the case of a pair of jets traveling through an expanding fluid, 
we showed that low momentum particles are dominant at large angles from the subleading jet.
These results indicate 
that the QGP-fluid+Jet model describes the CMS data \cite{Chatrchyan:2011sx} qualitatively
and importance of the QGP medium responses to the jet propagation.

\end{document}